\documentclass{article}
\usepackage{amsmath}
\usepackage{amssymb}
\newcommand{\be}{\begin{equation}}
\newcommand{\ee}{\end{equation}}
\begin{document}
\baselineskip18pt
 
\title{Quantum Potential and  Random Phase-Space Dynamics}
\author{Rados\l aw Czopnik \\
 Institute of Theoretical Physics, University of Wroc\l aw,\\
 PL - 50 209 Wroc\l aw, Poland\\
 and\\
 Piotr Garbaczewski\\
 Institute  of Physics, University of Zielona G\'{o}ra,\\
 PL - 65 069 Zielona G\'{o}ra, Poland}
 
\maketitle
 
\begin{abstract}
We  analyze limitations upon any kinetic theory  inspired
derivation of a probabilistic counterpart of the Schr\"{o}dinger
picture quantum dynamics. Neither dissipative nor  non-dissipative
stochastic phase-space processes based on the white-noise
(Brownian motion) kinetics are valid candidates unless  additional
  constraints (a suitable form of the energy
conservation law) are properly incorporated in the formalism.
\end{abstract}
 
PACS number(s): 02.50.-r, 05.40.-a, 03.65.-w

Born's statistical interpretation postulate in  quantum theory
may be viewed as a justification  for seeking   stochastic
counterparts of the Schr\"{o}dinger picture quantum dynamics, see
e.g.  \cite{nelson} - \cite{wallstrom} and references there-in.
The pertinent random dynamics typically has been introduced and
analyzed in the configuration space of the system, in terms of
Markovian diffusion-type processes which are compatible both with
the time evolution of the probability density $\rho = |\psi |^2$
and that of the wave function phase  $S$ in the Madelung
decomposition $\psi = \rho ^{1/2} exp(iS)$, the sigle-valuedness
condition being implicit, \cite{wallstrom}.
 
 In the corresponding hydrodynamical description of the Schr\"{o}dinger
  dynamics, we encounter two local conservation laws:
 the continuity equation
$\partial _t\rho = - \overrightarrow{\nabla }\cdot
(\overrightarrow{v} \rho )$  for the probability density $\rho $
and the Euler - type equation, controlling the space-time
dependence of the current velocity
$\overrightarrow{v}(\overrightarrow{x},t) = {\frac{\hbar }m}
\overrightarrow{\nabla } S(\overrightarrow{x},y)$ (a
gradient of the modified Hamilton-Jacobi  equation for $S$):
\begin{equation} (\partial _t + \overrightarrow{v} \cdot
\overrightarrow{\nabla }) \overrightarrow{v} = {\frac{1}m}
\overrightarrow{F} - \overrightarrow{\nabla } Q_q
\end{equation}
 where $Q_q= - {\frac{\hbar
^2}{2m^2}}{\frac{\Delta \rho ^{1/2}}{\rho ^{1/2}}}$  is the
familiar de Broglie - Bohm quantum potential, \cite{hol}. By
$\overrightarrow{F}$ we indicate the external force field acting
upon particles which in the conservative case coincides with
$-\overrightarrow{\nabla } V$ for a suitable potential
$V(\overrightarrow{x})$,  while the non-conservative case in our
further discussion will be restricted to the Lorentz force example
$\overrightarrow{F} = e (\overrightarrow{E} + \overrightarrow{v}
\times \overrightarrow{B})$ for charge $e$.
 
The de Broglie - Bohm potential $Q_q$ is usually interpreted to
have a pure quantum origin, hence one may be tempted to lend  a
status of an amusing curiosity  to the fact that pressure-type
potentials of the very same (de Broglie -Bohm) functional  form
notoriously appear in the so-called moment equations (local
conservation laws) associated with certain  kinetic partial
differential equations, like e.g. the Fokker - Planck - Kramers
equation for the phase-space random transport.

Indeed, hydrodynamical conservation laws for dissipative and
non-dissipative stochastic phase - space processes   give rise to
Euler-type and respective Hamilton-Jacobi equations where the
pertinent   contribution appears in the characteristic form $\pm
\overrightarrow{\nabla } \left[ 2d^2(t) \frac{\Delta \rho
^{1/2}}{\rho ^{1/2}}\right]$, \cite{czopnik02}.

In particular, the free Brownian motion is known
\cite{geilikman,gar92} to induce the current velocity
$\overrightarrow{v}= -  D \frac{\overrightarrow{\nabla }\rho }
\rho $ which obeys the continuity equation (that trivially yields
$\partial _t\rho = D\Delta \rho $) and  the local momentum
conservation law
\begin{equation}
(\partial _t + \overrightarrow{v}\cdot \overrightarrow{\nabla
})\overrightarrow{v} = - 2D^2 \overrightarrow{\nabla }
\frac{{\Delta  }\rho ^{1/2}}{\rho ^{1/2}}
\end{equation}
 where $D$ is the diffusion constant  (set formally $D\doteq \hbar/2m$
 for comparison with Eq. (1)). Notice the negative
sign on the right-hand-side of that equation, to be compared with
the corresponding quantum mechanical law.
 
This deceivingly   minor sign issue locates  the standard Brownian
motion in the framework of so-called Euclidean quantum mechanics
\cite{zambrini}, and stays behind the "Brownian recoil principle"
idea of Refs. \cite{vigier,gar99}, which was a proposed  recipe to
circumvent the breakdown of microscopic momentum and energy
conservation laws (a typical though not much disputed feature of
the Brownian motion).
 
Inspired by  the recent paper \cite{kaniadakis} let us consider
those issues from a broader kinetic theory perspective. We
investigate circumstances (including  various  constraints) under
which the classical phase-space kinetics can  approximate and
eventually may become  transformed into a stochastic counterpart
of the Schr\"{o}dinger picture  quantum dynamics.
 
To this end we need to analyze phase-space probability densities
$f(\overrightarrow{x},\overrightarrow{u},t)$ whose dynamics is
governed by the kinetic equation of the form
\begin{equation}
\left( \partial_t  +  \overrightarrow{u}\cdot
\overrightarrow{\nabla }_{\overrightarrow{x}} + {\frac{1}m}
\overrightarrow{F}\cdot \overrightarrow{\nabla
}_{\overrightarrow{u}} \right) f = C(f)
\end{equation}
where $m$ is a mass parameter, and the  term $C(f)$, in analogy
with the standard kinetic theory reasoning \cite{klim,huang}, is a
substitute for the so-called collision integral and encompasses
all  details about the environment (surrounding medium) action
and reaction  upon the propagated particle.
 
Evaluation of moment equations (local conservation laws)
associated with  the  kinetic equation (3)  relies  on the
concrete choice of  $C(f)$ and especially on the  microscopically
conserved quantities. In particular, the mass (probability)
conservation directly  follows from $\int C(f) d^3u =0$.  The
system total momentum is known  to be conserved in the force-free
case, only if $\int \overrightarrow{u} C(f)d^3u =
\overrightarrow{0}$.

To elucidate the particular role of those "collision invariants"
 let us consider  specific substitutes for   the
 collision integral,  which  impose  limitations on the number of
 respected microscopic conservation rules.\\

{\bf Case 1: $\int C(f) d^3u = 0$ only}\\
 
 Let us consider the \it standard \rm Brownian
motion in external conservative force fields,
\cite{chandra}.  We know a priori that  the external noise
intensity is determined by a parameter $q=D\beta ^{2}$ where
$D=\frac{kT}{m\beta }$, while the friction parameter $\beta $  is
given  by the Stokes formula $ m\beta =6\pi \eta a$. Consequently,
the effect of the surrounding medium on the motion of the particle
is described by two parameters: friction constant $\beta $ and
bath temperature $T$. Assumptions about the asymptotic
(equilibrium) Maxwell-Boltzmann distribution and the fluid
reaction upon the moving particle are here implicit.
 
The resulting (Markov)  phase - space diffusion  process is
completely determined by the transition probability density
$P\left( \left. \overrightarrow{x},\overrightarrow{u},t\right|
\overrightarrow{x}_{0},%
\overrightarrow{u}_{0},t_{0}\right) $,  which is typically expected
to be a fundamental solution of the Kramers equation, so that an initially given
phase-space  distribution
$f(\overrightarrow{x}_0,\overrightarrow{u}_0,t_0)$ is propagated according to:
 
\begin{equation}
\left( \partial _t + \overrightarrow{u}\cdot
\overrightarrow{\nabla }_{\overrightarrow{x}} +
{\frac{\overrightarrow{F}}{m}}\cdot  \overrightarrow{\nabla }
_{\overrightarrow{u}}
 \right)
f = C(f) =  \left( q\nabla _{\overrightarrow{u}}^{2}  + \beta
\overrightarrow{u} \cdot \overrightarrow{\nabla
}_{\overrightarrow{u}} \right) f
\end{equation}

First of all let us notice that in the present case $\int C(f) d^3u = 0$, while
$\int \overrightarrow{u} C(f) d^3u \neq \overrightarrow{0}$.
 
Accordingly, the continuity equation holds true for the marginal (spatial) probability
density $\rho = \int f d^3u $ while $ {\frac{1}\rho }\int \overrightarrow{u} C(f) d^3u =
- \beta \overrightarrow{v}(\overrightarrow{x},t)$ where $ \overrightarrow{v} \doteq
{\frac{1}\rho } \int \overrightarrow{u} f d^3u$. That has a devastating effect on the
form of the corresponding moment equation.

  The large friction (Smoluchowski) regime of the above
phase-space random dynamics is a classic \cite{chandra,nelson}. In
fact, by following the traditional pattern of the  hydrodynamical
formalism, \cite{klim,huang}, we  easily infer the closed system
of two (which is special to Markovian diffusions !) local
conservation laws for the Smoluchowski process  taking place in
the configuration space, \cite{gar99,gar00}:
\begin{eqnarray}
\partial _{t}\rho + \overrightarrow{\nabla }\cdot \left(
\overrightarrow{v} \rho \right)%
 &=&0 \\
( \partial _{t} + \overrightarrow{v}
\cdot \overrightarrow{\nabla }) \overrightarrow{v}%
 &=& \overrightarrow{\nabla }
\left( \Omega -Q\right)
\end{eqnarray}
where $\overrightarrow{v} \doteq
\overrightarrow{v}(\overrightarrow{x},t) = \frac{%
\overrightarrow{F}}{\ m\beta }- D\frac{\overrightarrow{\nabla }
\rho }{\rho }$
 defines the  so-called current velocity of Brownian particles and,
 when inserted to the continuity equation   allows to obtain  the
 Fokker-Planck equation, \cite{nelson}.
Here, the volume force (notice the positive
sign) $ + \overrightarrow{\nabla }\Omega $ instead of
$-\overrightarrow{\nabla }V$ as should have been to comply with  Eq. (1), reads:
\begin{equation}
\Omega =\frac{1}{2}\left( \frac{\overrightarrow{F}}{\ m\beta
}\right) ^{2}+D\overrightarrow{\nabla }\cdot \left(
\frac{\overrightarrow{F}}{\ m\beta }\right)
\end{equation}
while  the pressure-type contribution $ - \overrightarrow{\nabla }Q$
explicitly involves, \cite{gar99} (see also \cite{geilikman,gar92})
\begin{equation}
Q=2D^{2}\frac{\Delta \rho ^{1/2}}{\rho ^{1/2}}
\end{equation}
where  $\Delta  = \overrightarrow{\nabla }^2$ is the Laplace
operator.
 
 Eq. (6)  plays the role of the local momentum conservation law in the formalism.
 Let us however recall that $\overrightarrow{u}$ was \it not \rm a microscopic
 "collision invariant"  of the system.  This fact, when  combined with  the
 large friction regime, enforces a
marked difference in the local momentum conservation law in
comparison with the standard  Euler equation for a  nonviscous
fluid or gas:
\begin{equation}
 (\partial _t + \overrightarrow{v} \cdot
\overrightarrow{\nabla }) \overrightarrow{v}  =
\frac{\overrightarrow{F}}{\ m}\,  -  \,
\frac{\overrightarrow{\nabla } P} \rho
 \end{equation}
where $P(\overrightarrow{x})$ stands for the   pressure function
(to be fixed by a suitable equation of state) and
$\overrightarrow{F}$ is the very same (conservative
$-\overrightarrow{\nabla  }V$) force acting upon particles  as
that appearing in the   Kramers equation (3).
 
 To have a glimpse of a dramatic  difference between  physical
 messages conveyed respectively  by equations (6) and (9), it is
 enough to insert in (9) the standard equation of state
 $P(\overrightarrow{x}) = \alpha
\rho ^{\beta }$ with $\alpha , \beta > 0$ and choose
$\overrightarrow{F} = - \omega ^2 \overrightarrow{x}$ to represent
the harmonic attraction in Eqs. (2) - (9), see also \cite{gar99}.

{\bf Comment 1:}
In view of the breakdown of microscopic momentum and energy conservation laws in the
considered dissipative Brownian framework, one  needs supplementary procedures that
would compensate
those failures on the level of local conservation laws.
Markovian diffusion processes with the inverted sign of
$\overrightarrow{\nabla }(\Omega - Q)$ in the local momentum
conservation law (6) i. e. respecting
 
\begin{eqnarray}
( \partial _{t} + \overrightarrow{v}
\cdot \overrightarrow{\nabla }) \overrightarrow{v}%
 &=& \overrightarrow{\nabla }
\left( Q -\Omega \right)
\end{eqnarray}

instead of Eq. (6), were considered  in Ref. \cite{gar99} as
implementations of the "third Newton law in the mean" and could have been related to the
Schr\"{o}dinger-type dynamics according to:
\begin{equation}
i\partial _t \psi = - D\Delta \psi + {\frac{1}{2mD}}\Omega \psi \,
,
\end{equation}
 see e.g. Refs. \cite{gar99,gar00}.
Nonetheless, also under those premises,  the  volume force term
$-\overrightarrow{\nabla }\Omega$ in Eq.  (11) does \it not \rm in
general  coincide with the externally acting conservative force
contribution  (e.g. acceleration) ${\frac{1} m}\overrightarrow{F}
= - {\frac{1} m} \overrightarrow{\nabla } V$ akin to Eqs. (9) or (1).

{\bf Comment 2:}  Effects of  external force fields acting upon
particles  are significantly distorted while passing to the local
conservation laws in the large friction (Smoluchowski) regime.
That becomes   even more conspicuous  in case of the Brownian
motion of a charged  particle in the constant magnetic field. In
the Smoluchowski  (large friction) regime, friction completely
smoothes out any rotational (due to the Lorentz force) features of
the process. In the corresponding local momentum conservation law
there is \it no \rm volume force contribution  at all and merely
the "pressure-type" potential $Q$ appears  in a rescaled  form,
\cite{czopnik}:
\begin{equation}
Q = {\frac{\beta ^2}{\beta ^2 + \omega _c^2}} \cdot  2D^2
\frac{\Delta \rho ^{1/2}}{\rho ^{1/2}}
\end{equation}
where $\beta $ is the (large) friction parameter and $\omega _c =
\frac{e B}{m}$ is the rotational  frequency of the charge $q_e$
particle in a constant homogeneous  magnetic field
$\overrightarrow{B}= (0,0,B)$. Clearly, for moderate  frequency
values $\omega _c$ (hence the magnetic field intensity) and
sufficiently   large $\beta $ even this minor scaling remnant of
the original Lorentz force would  effectively disappear, yielding
the free Brownian dynamics (2).\\

The last  observation should  be compared  with results of  Refs.
\cite{newman} where  frictionless  stochastic processes were
invoked to analyze  situations present in magnetospheric
environments. Specifically, one deals there with charged particles
in a  locally uniform magnetic field  which experience stochastic
electrical forcing. In the absence of friction, the rotational
Lorentz force  input should clearly  survive  when passing to the
local conservation laws, in plain contrast with the Smoluchowski
regime. By disregarding friction  it is also possible to reproduce
exactly the conservative external force acting upon particles in
the local conservation laws, as originally suggested by
\cite{nelson}.\\

{\bf Case 2: $\int C(f)d^3u =0$ and $\int \overrightarrow{u} C(f) d^3u
= \overrightarrow{0}$} \\
 
Let us consider the frictionless phase - space dynamics
in some detail. Clearly,  that corresponds to dropping the frictional
 ($\beta $-dependent) contribution in the right-hand-side of Eq. (4).
 We immediately realize that the previous obstacle pertaining to the
 "collision invariant" $\overrightarrow{u}$ disappears. Indeed, now
 $\int \overrightarrow{u} C(f)d^3u = \overrightarrow{0}$.
 
Let us discuss the  non-dissipative  random dynamics in two phase-space dimensions
instead of six. The motion is governed  by a transition
  density
$P\left( \left. x,u,t\right| x_{0},u_{0},t_{0}\right) $ which
uniquely defines the  corresponding  time homogeneous phase-space
Markovian diffusion process.
Here, the function $P$ is the fundamental solution of  the
Fokker-Planck equation:
\begin{equation}
\frac{\partial P}{\partial t}=-u\frac{\partial P}{\partial x}+q\frac{%
\partial ^{2}P}{\partial u^{2}}
\end{equation}
in the form   first given  by Kolmogorov \cite{kolmogorov}:
\begin{equation}
P\left( \left. x,u,t\right| x_{0},u_{0},t_{0}=0\right) =
\frac{1}{2\pi }\frac{%
\sqrt{12}}{2qt^{2}}\exp \left[ -\frac{\left( u-u_{0}\right) ^{2}}{4qt}-\frac{%
3\left( x-x_{0}-\frac{u+u_{0}}{2}t\right) ^{2}}{qt^{3}}\right]\, .
\end{equation}
 
Here $q$ stands for the noise intensity parameter which  may take
an arbitrary non-negative  value (in contrast to the dissipative
case where fluctuation-dissipation relations set a connection of
$q$ with $\beta $ and $T$).
 
We are interested in passing to a hydrodynamical picture,
following the traditional recipes \cite{klim,huang}. To this  end
we need to propagate certain initial probability density  and
investigate effects of the random dynamics. Let us
 choose most obvious (call it natural) example of:
 
\begin{equation}
f_{0}\left( x,u\right) =\left( \frac{1}{2\pi a^{2}}\right) ^{\frac{1}{2}%
}\exp \left( -\frac{\left( x-x_{ini}\right) ^{2}}{2a^{2}}\right)
\left( \frac{1}{2\pi b^{2}}\right) ^{\frac{1}{2}}\exp \left(
-\frac{\left( u-u_{ini}\right) ^{2}}{2b^{2}}\right)\, .
\end{equation}
so that  at time $t$ we have $f\left( x,u,t\right) =\int
P\left( \left. x,u,t\right| x_{0},u_{0},t_{0}=0\right)
f_{0}\left( x_{0},u_{0}\right) dx_{0}du_{0}$.
 
Since $P\left( \left. x,u,t\right| x_{0},u_{0},t_{0}\right) $ is
the fundamental solution of the Kramers equation,  the joint
density $f\left( x,u,t\right) $ is also the solution and  can
be written in  the familiar, \cite{chandra},  form  of
 
\begin{equation}
W\left( R,S\right) =\left( \frac{1}{4\pi ^{2}\left(
eg-h^{2}\right) }\right) ^{\frac{1}{2}}\exp \left[
-\frac{gR^{2}-2hRS+eS^{2}}{2\left( eg-h^{2}\right) }\right]
\end{equation}
 for $f(x,u,t) = W(R,S)$. However, in the present case
functional entries are adopted to the frictionless motion and read
as follows:
 $$S=u-u_{ini}$$
 $$R=x-x_{ini}-u_{ini}t$$
 $$e=a^{2}+b^{2}t^{2}+\frac{2%
}{3}qt^{3}$$
 \begin{equation}
 g=b^{2}+2qt
\end{equation}
$$h=b^{2}t+qt^{2}\, .$$
 
The marginals $\rho \left( x,t\right) =\int f\left(
x,u,t\right) du$ and $\rho \left( u,t\right) =\int f\left(
x,u,t\right) dx$\ are
 
\begin{equation}
\rho \left( x,t\right) =\left( \frac{1}{2\pi f}\right)
^{\frac{1}{2}}\exp \left( -\frac{R^{2}}{2f}\right) =\left(
\frac{1}{2\pi \left( a^{2}+b^{2}t^{2}+\frac{2}{3}qt^{3}\right)
}\right) ^{\frac{1}{2}}\exp \left(
-\frac{\left( x-x_{ini}-u_{ini}t\right) ^{2}}{2\left( a^{2}+b^{2}t^{2}+\frac{%
2}{3}qt^{3}\right) }\right)
\end{equation}
and
 
\begin{equation}
\rho \left( u,t\right) =\left( \frac{1}{2\pi g}\right)
^{\frac{1}{2}}\exp \left( -\frac{S^{2}}{2g}\right) =\left(
\frac{1}{2\pi \left( b^{2}+2qt\right) }\right) ^{\frac{1}{2}}\exp
\left( -\frac{\left( u-u_{ini}\right) ^{2}}{2\left(
b^{2}+2qt\right) }\right)
\end{equation}
 
Let us introduce an auxiliary (reduced) distribution
 
\begin{equation}
\widetilde{W}\left( S|R\right) =\frac{W\left( S,R\right) }{\int
W\left( S,R\right) dS}=\left( \frac{1}{2\pi \left(
g-\frac{h^{2}}{e}\right) }\right) ^{\frac{1}{2}}\exp \left(
-\frac{\left| S-\frac{h}{e}R\right| ^{2}}{2\left(
g-\frac{h^{2}}{e}\right) }\right)  \label{aux}
\end{equation}
where in the denominator we recognize    the marginal spatial
distribution $\int W(S,R)dS = \rho $.
 
Following the standard hydrodynamical picture method
\cite{klim,huang,gar00} we define local (configuration space
conditioned) moments: $\left\langle u\right\rangle _{x}=\int
u\widetilde{W}du$ and $\left\langle u^{2}\right\rangle _{x}=\int
u^{2}\widetilde{W}du$. From (\ref{aux}) it follows that
 
\begin{equation}
\left\langle u\right\rangle _{x} =  u_{ini}+\frac{h}{e}R
=u_{ini}+\frac{b^{2}t+qt^{2}}{a^{2}+b^{2}t^{2}+\frac{2}{3}qt^{3}}\left[
x-x_{ini}-u_{ini}t\right]
\end{equation}
 
\begin{equation}
\left\langle u^{2}\right\rangle _{x}-\left\langle u\right\rangle
_{x}^{2}=\left( g-\frac{h^{2}}{e}\right) =\frac{q\,t^{3}\,\left(
2\,b^{2}+q\,t\right) +3\,a^{2}\,\left( b^{2}+2\,q\,t\right) }{%
3\,a^{2}+t^{2}\,\left( 3\,b^{2}+2\,q\,t\right) }
\end{equation}
 
The continuity ($0$-th moment) and the momentum
conservation (first moment) equations come out in the form
 
\begin{eqnarray}
\frac{\partial \rho }{\partial t}+\frac{\partial }{\partial x}\left(
\left\langle u\right\rangle _{x}\rho \right) &=&0 \\
\frac{\partial }{\partial t}\left( \left\langle u\right\rangle _{x}\rho \right) +%
\frac{\partial }{\partial x}\left( \left\langle u^{2}\right\rangle
_{x}\rho \right) &=&0\, .
\end{eqnarray}
 
These equations yield the  local momentum conservation law in the
 form (set $v(x,t) \doteq  <u>_x$:
\begin{equation}
\left( \frac{\partial }{\partial t}+\left\langle u\right\rangle _{x}\frac{%
\partial }{\partial x}\right) \left\langle u\right\rangle _{x}=-\frac{1}{w}%
\frac{\partial P_{kin}}{\partial x}  \label{lmcl_1}
\end{equation}
where we  encounter the standard \cite{huang} textbook  notion of
the pressure function
\begin{equation}
P_{kin}\left( x,t\right) =\left[ \left\langle u^{2}\right\rangle
_{x}-\left\langle u\right\rangle _{x}^{2}\right] w\left(
x,t\right) \, .
\end{equation}
 
The marginal density  $\rho $ obeys  $\frac{\nabla \rho }{\rho }=-2e\nabla %
\left[ \frac{\Delta \rho ^{1/2}}{\rho ^{1/2}}\right] $ and that in turn
implies
 
\begin{equation}
-\frac{1}{w}\frac{\partial P_{kin}}{\partial x}=2\left(
eg-h^{2}\right) \nabla \left[ \frac{\Delta
\rho ^{1/2}}{\rho ^{1/2}}\right]\, .
\end{equation}
 
As a consequence, the local conservation law  takes
the form (we ultimately set $<u>_x \doteq v$):
 
\begin{equation}
\left( \frac{\partial }{\partial t}+ v\cdot \nabla \right) v= -\frac{\nabla
P_{kin}}{\rho }= + 2\left( eg-h^{2}\right) \nabla \left[ \frac{\Delta
\rho ^{1/2}}{\rho ^{1/2}}\right] \doteq + \nabla Q
\end{equation}
where (we point out the plus   sign in the above, see e.g.  Eq.
(10))
 
\begin{equation}
eg-h^{2}=a^{2}b^{2}+2a^{2}qt+\frac{2}{3}b^{2}qt^{3}+
\frac{1}{3}q^{2}t^{4} \doteq D^2(t)
\end{equation}
and  by adopting the notation $D^2(t) \doteq  eg - h^2$  we get
$-{\frac{1}\rho } {\frac{\partial P_{kin}}{\partial x}} = +
\overrightarrow{\nabla } Q$ with the functional form of
$Q(\overrightarrow{x},t)$ given by Eq. (8). Here,  instead of a
diffusion constant $D$ we insert  the (positive) time -dependent
function $D(t)$.
 
With those notational adjustments, we recognize in Eq. (28)  a
consistent Euler form of the local momentum conservation law,  in
case of vanishing volume forces (c.f. Eqs. (6), (8), (10) for
comparison).
 
A carefully executed, tedious calculation allows to demonstrate,
\cite{czopnik02}, that  an analogous result holds true in case of
a harmonic attraction and for a nonconservative example of the
Lorentz force in action. Both  volume forces appear undistorted
(that was \it not \rm  the case in  the large friction regime)  in
the corresponding  local momentum conservation laws. Indeed, we
recover a universal relationship:
 
\begin{equation}
\left[ \partial _{t}+ \overrightarrow{v} \cdot \overrightarrow{\nabla }\right]
\overrightarrow{v} = \frac{\overrightarrow{F}}m
+2d^2(t)
\overrightarrow{\nabla }\left[ \frac{%
\Delta \rho ^{1/2}}{\rho ^{1/2}}\right]
\end{equation}
where $\overrightarrow{F}$ denotes external force  acting on the
particle, and $d=\left( \det C\right) ^{\frac{1}{n}}$ where $C$ is
the
covariance matrix of random variables (vectors) $\overrightarrow{S}$ and $%
\overrightarrow{R}$ (defined for each system) and $n$ stands for
the dimension of configuration space of an  appropriate system.
 
\begin{enumerate}
\item  free particle: $\overrightarrow{R}=x$, $F\equiv 0$, $n=1$
 
\item  charged particle in a constant magnetic field: $\overrightarrow{R}%
=\left( x,y\right) $, $\overrightarrow{F}= e \left\langle
\overrightarrow{u}\right\rangle _{\overrightarrow{x}}\times \overrightarrow{B%
}$, $n=2$
 
\item  harmonically bound particle: $\overrightarrow{R}=x$, $F=- m \omega ^{2}x$%
, $n=1$
\end{enumerate}
 
In case of harmonic and magnetic confinement, we need to have
identified parameter range regimes  that  allow for a positivity
of the time dependent coefficient
 $d(t) \doteq D(t)$ (in the force-free case it is positive with no
 reservations), in  the pressure-type contribution  acquiring
 a characteristic form of
 $- {\frac{\overrightarrow{\nabla } \cdot \overleftrightarrow{P}} w} =
 + \overrightarrow{\nabla }Q $.
Indeed, only by means of a proper balance between $q$ and
$\omega_c$ we can achieve a positivity of  the  coefficient $d^2$ in case of
the charged particle in a magnetic field:
\begin{equation}
d^2(t) = eg-h^{2}-k^{2}=
\end{equation}
$$
a^{2}\,b^{2}\,-\frac{8\,q^{2}}{\omega _{c}^{4}}+\frac{%
4\,b^{2}\,q\,t}{\,\omega _{c}^{2}}+\frac{4\,q^{2}\,t^{2}}{\,\omega _{c}^{2}}%
+2\,a^{2}\,q\,t+\frac{8\,q^{2}}{\omega _{c}^{4}}\,\cos (t\,\omega _{c})-%
\frac{4\,b^{2}\,q}{\omega _{c}^{3}}\,\sin (t\,\omega _{c}) \, .
$$
 
The  time-dependent coefficient for the frictionless harmonic
attraction  reads:
\begin{equation}
d^2(t) = eg-h^{2}=
\end{equation}
$$
\frac{-q^{2}+2\,b^{2}\,q\,t\,\omega ^{2}+2\,q^{2}\,t^{2}\,\omega
^{2}+2\,a^{2}\,b^{2}\,\omega ^{4}+2\,a^{2}\,q\,t\,\omega
^{4}+q^{2}\,\cos (2\,t\,\omega )+q\,\omega \left(
-b^{2}+a^{2}\,\omega ^{2}\right) \,\sin (2\,t\,\omega )}{2\,\omega
^{4}}
$$
and a proper balance between $q$  and $\omega $ needs to be
maintained again to secure a positivity of $d^2(t)$.
 
In Ref. \cite{czopnik02} we have investigated the above
expressions  in the low noise intensity regime and for rather
short duration times of the pertinent stochastic processes.  That
was motivated by the major  conceptual input of
\cite{newman,czopnik02} that undamped random flights in external
force fields may have physical relevance when dissipative time
scales  are much longer than the time duration of processes of
interest, including the  particle life-time.
Under those premises, we can view the noise intensity parameter
$q$ as a book-keeping label and investigate leading  terms in
 all expressions  encompassing  small $q$/short time$t$ effects
  in the  hitherto considered random dynamics.
In particular, it is obvious that by neglecting all $q$-dependent
terms we readily arrive at the leading term:  $d^2(t) \rightarrow
a^2b^2 \doteq D^2$. After adopting this  notation, Eq. (31) acquires a
conspicuous quantum mechanical form in the leading order of
$q$-dependent series expansion.
 
To elucidate the previous observation, let us explicitly associate
the  $q\ll 1$  local momentum conservation law with that known to
be appropriate for  quantum harmonic oscillator.
 
We impose the following condition on the dispersion parameters
$a=\sqrt{\left\langle x^{2}\right\rangle -\left\langle
x\right\rangle ^{2}}$ and $b=\sqrt{\left\langle v^{2}\right\rangle
-\left\langle v\right\rangle ^{2}}$ of initial distributions $\rho
_{0}\left( x\right) $ and $\rho _{0}\left( v\right) $ respectively
 
\begin{equation}
d^2(0) = a^{2}\,b^{2} \doteq    \left( \frac{\hbar }{2m}\right) ^{2}\,  .
\end{equation}
 
This condition is  obviously equivalent to imposing a priori the
Heisenberg uncertainty relation $a\left( mb\right)
=\sqrt{\left\langle x^{2}\right\rangle -\left\langle
x\right\rangle ^{2}}\sqrt{\left\langle mv^{2}\right\rangle
-\left\langle mv\right\rangle ^{2}}=\frac{\hbar }{2}$. Upon an
additional demand $ b^{2}=a^{2}\omega ^{2}$ which is an identity
for the choice of $b^{2} =\frac{\hbar \omega }{2m}$, $a^{2}
=\frac{\hbar }{2m\omega }$, we recover
\begin{equation}
\rho \left( x,t\right) =\left( \frac{m\omega }{\hbar \pi }%
\right) ^{\frac{1}{2}}\exp \left( -\frac{m\omega }{\hbar }\left(
x-x_{ini}\cos \omega t\right) ^{2}\right)
\end{equation}
which directly corresponds  to the quantum evolution of the
coherent state of the harmonic oscillator.
 
Let us point out  that the Planck constant entered the formalism
through an \it additional  \rm assumption about an initial form  (widths)
of the phase-space
probability density, see Eq. (33).

To conclude, we have demonstrated that  neither  large friction
nor frictionless  cases may be regarded as  valid  probabilistic
phase-space motion ancestors for a  consistent stochastic
counterpart of the quantum Schr\"{o}dinger picture dynamics.
 
 The formal reason for those failures is rooted in violations of the
microscopic conservation laws  as expressed through the
non-vanishing "collision integrals". However,  the frictionless
case appears to be sufficiently close (at least  in the small $q$
and short duration time $t$ regime) to the quantum dynamics in its
hydrodynamical description. Large friction dynamics, even if
augmented by the concept of the "Brownian recoil principle",
\cite{vigier}, is incapable of reproducing  correctly the external
force effects in the local momentum conservation laws
 
An interesting point, worth further exploration, is that in  the
frictional case    \it one \rm  only  "collision invariant" has
been respected, while in case of the frictionless dynamics, \it
two \rm basic "collision invariants" were in usage. It is rather
obvious that in case of the  Brownian motion inspired kinetics,
 the "collision integral" $\int \overrightarrow{u}^2 C(f)d^3u$ is
non-vanishing. Therefore, any kinetic framework  exploiting
$\overrightarrow{u}^2$ or its analog (cf. \cite{huang}) as a
microscopically conserved quantity (that is the case in Ref.
\cite{kaniadakis}, albeit this restriction  is not used at all in
major derivations) must definitely depart form the "plain" white
noise scenario. The usage of additional
 constraints  that are capable of compensating  violations  of
 the microscopic energy conservation law, is unavoidable.

\end{document}